\documentclass[journal,11pt, onecolumn]{IEEEtran}
\usepackage{times,amssymb,amsmath,epsfig,nicefrac,euscript,cite,mathrsfs,mathrsfs}
\usepackage{hyperref}
\usepackage{color}
\newcommand\nc\newcommand
\nc\bfa{{\boldsymbol a}}\nc\bfA{{\boldsymbol A}}\nc\cA{{\mathcal A}}
\nc\bfb{{\boldsymbol b}}\nc\bfB{{\boldsymbol B}}\nc\cB{{\mathcal B}}
\nc\bfc{{\boldsymbol c}}\nc\bfC{{\boldsymbol C}}\nc\cC{{\mathcal C}}
\nc\sC{{\mathscr C}}
\nc\bfd{{\boldsymbol d}}\nc\bfD{{\boldsymbol D}}\nc\cD{{\mathcal D}}
\nc\bfe{{\boldsymbol e}}\nc\bfE{{\boldsymbol E}}\nc\cE{{\mathcal E}}
\nc\bff{{\boldsymbol f}}\nc\bfF{{\boldsymbol F}}\nc\cF{{\mathcal F}}
\nc\bfg{{\boldsymbol g}}\nc\bfG{{\boldsymbol G}}\nc\cG{{\mathcal G}}
\nc\bfh{{\boldsymbol h}}\nc\bfH{{\boldsymbol H}}\nc\cH{{\mathcal H}}
\nc\bfi{{\boldsymbol i}}\nc\bfI{{\boldsymbol I}}\nc\cI{{\mathcal I}}
\nc\bfj{{\boldsymbol j}}\nc\bfJ{{\boldsymbol J}}\nc\cJ{{\mathcal J}}
\nc\bfk{{\boldsymbol k}}\nc\bfK{{\boldsymbol K}}\nc\cK{{\mathcal K}}
\nc\bfl{{\boldsymbol l}}\nc\bfL{{\boldsymbol L}}\nc\cL{{\mathcal L}}
\nc\bfm{{\boldsymbol m}}\nc\bfM{{\boldsymbol M}}\nc\sM{{\mathscr M}}
\nc\bfn{{\boldsymbol n}}\nc\bfN{{\boldsymbol N}}\nc\cN{{\mathcal N}}
\nc\bfo{{\boldsymbol o}}\nc\bfO{{\boldsymbol O}}\nc\cO{{\mathcal O}}
\nc\bfp{{\boldsymbol p}}\nc\bfP{{\boldsymbol P}}\nc\cP{{\mathcal P}}
\nc\bfq{{\boldsymbol q}}\nc\bfQ{{\boldsymbol Q}}\nc\cQ{{\mathcal Q}}
\nc\bfr{{\boldsymbol r}}\nc\bfR{{\boldsymbol R}}\nc\cR{{\mathcal R}}
\nc\bfs{{\boldsymbol s}}\nc\bfS{{\boldsymbol S}}\nc\cS{{\mathcal S}}
\nc\bft{{\boldsymbol t}}\nc\bfT{{\boldsymbol T}}\nc\cT{{\mathcal T}}
\nc\bfu{{\boldsymbol u}}\nc\bfU{{\boldsymbol U}}\nc\cU{{\mathcal U}}
\nc\bfv{{\boldsymbol v}}\nc\bfV{{\boldsymbol V}}\nc\cV{{\mathcal V}}
\nc\bfw{{\boldsymbol w}}\nc\bfW{{\boldsymbol W}}\nc\cW{{\mathcal W}}
\nc\bfx{{\boldsymbol x}}\nc\bfX{{\boldsymbol X}}\nc\cX{{\mathcal X}}
\nc\bfy{{\boldsymbol y}}\nc\bfY{{\boldsymbol Y}}\nc\cY{{\mathcal Y}}
\nc\bfz{{\boldsymbol z}}\nc\bfZ{{\boldsymbol Z}}\nc\cZ{{\mathcal Z}}

\def\Var{\qopname\relax{no}{Var}}

\def\avg{{\mathbb E}}

\def\supp{\qopname\relax{no}{supp}}
\newtheorem{theorem}{Theorem}
\newtheorem{definition}{Definition}
\newtheorem{lemma}[theorem]{Lemma}
\newtheorem{proposition}[theorem]{Proposition}
\newtheorem{corollary}[theorem]{Corollary}

\newtheorem{remark}{\indent Remark}

\newcommand\ff{{\mathbb F}}

\newcommand{\remove}[1]{}
\begin{document}
\allowdisplaybreaks
\title{Group testing schemes from codes and designs}

\author{Alexander Barg and  Arya Mazumdar

 
 \thanks{A.~ Barg is with the Department of ECE and ISR, University of Maryland, College Park, MD 20742 and Institute for
Problems for Information Transmission, Russian Academy of Sciences, Moscow, Russia. Email: 
abarg@umd.edu. Research supported by NSF grants CCF1217894, CCF1217245, CCF1422955, and CCF1618603.}

 \thanks{A.~Mazumdar  is with the College of Information and Computer Sciences, University of Massachusetts Amherst, Amherst, MA 01003. Email: arya@cs.umass.edu.
 Research supported in part by NSF grants  CCF1318093, CCF1642658, CCF 1642550 and CCF 1618512.} 
  
}

\maketitle

\begin{abstract} 
In group testing, simple binary-output tests are designed to identify a small number $t$ of defective items  
that are present in a large population of $N$ items. Each test takes as input a group of items and produces
a binary output indicating whether the group is free of the defective items or contains one or more of them.

In this paper we study a relaxation of the combinatorial group testing problem. 
A  matrix is called $(t,\epsilon)$-disjunct if it gives rise to a nonadaptive group testing scheme with the property of identifying a uniformly random $t$-set of defective subjects out of a population of size $N$ with false positive probability of an item at most $\epsilon$. 
We establish a new connection between $(t,\epsilon)$-disjunct matrices and error correcting codes based on the dual distance of the codes
and derive estimates of the parameters of codes that give rise to such schemes. 
Our methods rely on the moments of the distance distribution of codes and inequalities for moments of sums of independent
random variables. We also provide a new connection between group testing schemes
and combinatorial designs. 
\end{abstract}


\section{Introduction}

Suppose that the elements of a finite population of size $N$ contain a small number of defective elements. The elements
are tested in groups, and the collection of tests is said to form a group testing scheme if the outcomes of the tests
enable one to identify any defective configuration size at most $t$. 
Let the number of tests in a group testing scheme be $M$. Then
constructing a non-adaptive group testing scheme is equivalent to constructing a binary {\em test matrix}
of dimensions $M \times N$ where the $(i,j)$-th entry is $1$ if the $i$th test includes the $j$th element and
is $0$ otherwise. Each row of the matrix corresponds to a test, and the result of this test is positive
if the indices of ones in the row have a nonempty intersection with the indices of the defective configuration.
The smallest possible number of tests in terms of the total number of subjects $N$ and the maximum number of
defective elements $t$ is known to satisfy
 $ M = \Omega\big(\frac{t^2}{\log{t}} \log N\big);
 $ 
see \cite{d1982bounds,ruszinko94,furedi96}.

A construction of group testing schemes using error-correcting codes and code concatenation
appeared in the foundational paper by Kautz and Singleton \cite{kautz1964nonrandom}.
They introduced a two-level construction in which a $q$-ary ($q>2$) Reed-Solomon code is concatenated
with a unit-weight binary code. The resulting vectors are used as columns of the testing matrix. Since every symbol of the Reed-Solomon codeword is replaced by
a binary vector of Hamming weight one, the overall code is formed of codewords of a fixed Hamming weight. Overall, with $M = O(t^2 \log^2 N)$ tests this scheme
identifies any defective configuration of size up to $t$. More importantly, this construction offered a general method of obtaining
test matrices from error-correcting codes.
Many later constructions of group testing schemes also rely on Reed-Solomon codes and code concatenations; among them 
\cite{dyachkov00,sid09,hwang1987non,nguen88}. An explicit constructions of non-adaptive group testing schemes with $M = O(t^2 \log N)$ 
was presented in \cite{porat2008explicit}.

It has been suggested
to construct schemes that permit a small probability of error (i.e., allowing false positives).
Such schemes were considered under the name of {\em almost disjunct matrices } or {\em weakly separated designs} in 
\cite{malyutov1977mathematical,malyutov1978separating}, \cite{zhigljavsky2003probabilistic} and independently in \cite{macula2004trivial}.
With this relaxation it is possible to reduce the number of tests to be proportional to $t\log N$ \cite{zhigljavsky2003probabilistic}; however,
this result is not constructive. In terms of constructive results, a scheme with $t^{3/2}\sqrt{\log N}$ tests was presented in \cite{mazumdar2012almost}. The recent work of Gilbert et al. \cite{gilbert2012recovering} 
 suggests a way to construct weakly separated designs with $O(t \,{\rm  poly}(\log N))$ tests by partitioning the
 subjects into blocks of equal size and using optimal non-adaptive tests independently for each block. 
Finally, an explicit (non-probabilistic) construction of nonadaptive schemes with the 
number of tests proportional to  
$t \log^2 N/\log t$ was presented  in \cite{mazumdar2015}. All these works allow a small probability of existence of false positives.

The construction of Kautz and Singleton \cite{kautz1964nonrandom} and many others above are based on {constant weight error-correcting codes}. 
Estimates of the parameters of the group testing schemes from constant weight codes were obtained using the minimum
distance of the code \cite{kautz1964nonrandom} and more recently  using the average distance of the code \cite{mazumdar2012almost,mazumdar2015}. 

Our present work takes a different approach, relating construction of almost disjunct matrices and the {\em dual distance} of codes.
Our main contribution consists of a refined analysis of constructions of group testing schemes that relates the number of tests $M$ 
to the dual distance of the (constant weight) code and moments of the distance distribution. 

The main ideas of our analysis involve techniques from coding theory, algebraic combinatorics, and probability, and are as follows. We form a group testing matrix whose columns are the codewords of a binary
constant weight code $\cC$. A simple observation given in Prop.~\ref{prop:pks} connects the probability $P$ of violating the group testing recovery  condition to the distribution of 
distances in the code $\cC$  (the probability comes assuming the defective elements to be uniformly distributed). This enables us to transform the problem of estimating $P$ to the question of estimating moments of
the distance distribution of the code. 
In the next step, we consider examples of group testing schemes obtained from
several known families of nonbinary codes via the Kautz-Singleton map. 
We obtain explicit expressions for the moments of order that does not exceed
the dual distance of the nonbinary code (the strength of the combinatorial design formed by the code). Classical inequalities for the moments of sums of independent random variables give upper bounds on the
moments, resulting in upper estimates of the probability $P.$ 

Continuing the above line of thought, we turn to binary constant weight codes. We again connect the problem of estimating
the probability $P$ with a question about moments of the distance distribution of
constant weight codes. Using simple facts from the theory of association schemes, we find an exact expression for the
moments of order less than the dual distance of the code. Bounds for the probability $P$ are again obtained using classical 
inequalities from probability theory. We note that, in the case of constant weight codes, the dual distance is related to the strength of the design formed by the codewords, and so the problem of constructing good almost disjunct matrices
can be expressed in terms of combinatorial designs. This approach to the construction of group testing schemes does not
involve the Kautz-Singleton construction.

Apart from \cite{mazumdar2012almost,mazumdar2015}, the connection 
between error-correcting codes and weakly separated designs was known only for the very specific family of {\em maximum distance separable} codes \cite{macula2004trivial,bas13}, for which much more than the
dual distance is known. 

\subsection{Contributions of this paper} 
In this work we estimate parameters of almost disjunct matrices using the distance distribution of codes.

\subsubsection{Matrices from nonbinary codes} In this part, we show that group testing schemes from Reed-Solomon (RS) codes and codes on algebraic curves have the following parameters while allowing a small probability of false positives (that approaches zero with $N$ for RS codes and with the size of the code alphabet $q$ for the other cases):
  \begin{equation*}\begin{array}{llll}
  \text{RS codes} & & &M=O\big(\max\big\{t^2,t\big(\frac{\log N}{\log t}\big)^3\big\}\big)  \\
  \text{Hermitian codes ($q$ is a prime power)} &t=q^2 & N = q^{2q^2} &M=q^5 \\
  \text{Suzuki codes ($q$ is a power of 2)} &t=q^2& N= (2q)^{2q^3+q+1} & M=8q^6 
  \end{array}
  \end{equation*}
Note that in the case of RS codes, we directly give an expression for $M$ in terms of $t$ and $N$ rather than individual expressions. This can be done because the relation between $N,t,$ and $q$ is given in the form of an inequality (see \eqref{eq:Ntq} and the discussion after it below).
In the other two examples, the values of the parameters $N, t,$ and $M$ are uniquely determined by the value of $q$.


\subsubsection{Almost disjunct matrices from constant weight codes} Our main contribution in this part can be summarized as follows. We show that a constant weight code of dual distance $d'$ gives rise to 
a group testing scheme   that can, with $O(t (d')^2)$ tests, identify
all items in a random defective configuration of size  $t$  with probability of false positive for an element (outside of defective set) at most $\exp(-d')$  (see, Corollary \ref{cor:last}).

To be able to construct almost disjunct matrices, we turn to known results about the existence of constant weight codes with large dual distance (i.e., of combinatorial designs; see Def.~\ref{def:designs} and the discussion after it). This problem was
addressed in a recent paper \cite{kuperberg2013probabilistic}. According to it, there exist codes of length $M$, size $N,$ and dual distance $d'$ such that,
  \begin{equation}\label{eq:KLP}
d' = \Omega\Big(\frac{\log N}{\log (M/d')}\Big).
  \end{equation}
Relying on this result, we obtain group testing schemes with $O(t \log^2N /\log^2 t)$ tests that can identify all items in a random defective configuration of size  $t$  with probability of false positive for an element at most $\exp(-\log N/\log t);$ see Corollary \ref{cor:last1}. Comparing this with \cite{mazumdar2015}, while we improve on the number of tests by a factor of $\log t$, we obtain a slower decline of the probability of false positives, which was $O(1/{\rm poly}(N))$ therein.

A caveat about the result of \cite{kuperberg2013probabilistic} given in \eqref{eq:KLP} is that it is an existence claim rather than an explicit construction. However, we emphasize that the main new idea of our work is the connection between the dual distance of codes and group testing properties. Hence whenever constructions of combinatorial designs are available, we can use them as building blocks for 
constructing group testing scheme via the approach discussed in the paper. Indeed, recently we have found  \cite{UMB16} that constant weight codes formed by the fixed-weight codewords of BCH code perform consistently better compared to the random test matrices in terms of false positives, which gives empirical evidence that combinatorial designs are good candidates for nonadaptive group testing (some more details about \cite{UMB16} are given in the concluding Section~\ref{sect:outlook}).

\vspace*{.1in}
\subsection{Plan of the paper}
The paper is organized as follows. Some preliminaries about group testing and the mathematical tools that we use are provided
in Section~\ref{sec:defnot}. Group testing schemes from nonbinary codes with large dual distance are discussed in Section~\ref{sect:AG},
where the main result is Theorem~\ref{prop:codes}, and
schemes from binary constant weight codes (designs) are discussed in Section~\ref{sect:CW}. Here the main results are Theorem \ref{prop:prop10}   and Corollaries   \ref{cor:last} and \ref{cor:last1}.

\section{Preliminaries}\label{sec:defnot}
\subsection{Group testing schemes}
Define the {\em support} of a vector $\supp(x), x\in \ff_q^n$ as the set of coordinates  where $x$ has nonzero
entries. The support of a set of vectors $X=\{x_i, i\ge 1\}$ is the union of supports $\cup_{i\ge 1}\supp(x_i).$

\begin{definition}
An $M \times N$ binary matrix $A$ is called $t$-disjunct if the support of any of its columns
is not contained in the union of the supports of any other $t$ columns.
\end{definition}

It is easy to see that a $t$-disjunct matrix gives a group testing scheme
that identifies any defective set up to size $t$. Conversely,
any group testing scheme that identifies any defective set up to size $t$ must be a $(t-1)$-disjunct
matrix \cite{du1993combinatorial}. To a great advantage,  disjunct matrices support a simple identification
algorithm that runs in time $O(Nt).$ Indeed, any element that participates in a test with a negative outcome is not defective.
After we perform all the tests and weed out all the non-defective elements, the disjunctness property of the matrix
guarantees that all the remaining elements are defective.

A few words on notation. Let $[ N]:=\{1,2,\dots,N\}$ and let $\cP_t(N)$ denote the set of $t$-subsets of $[N].$ The usual notation 
for probability $\Pr$ is used to refer a probability measure which will be understood from the context. 
Separate notation will be used for some frequently encountered probability spaces.
In particular, we use $P_{R_t}$ to denote the uniform probability distribution on $\cP_t(N)$.
If we need to choose a random $t$-subset $I$ and a random index in $[N]\backslash I,$ we use the notation $P_{R_t'}$.

\vspace*{.05in}
\begin{definition}\label{def:adm}
For any $\epsilon >0$, an $M \times N$ matrix $A$ with columns $a_1,a_2,\dots, a_N$,  is called $(t,\epsilon)$-disjunct if  
\begin{align}
P_{R_t'}(\{I \in \cP_t(N), &\,\,   j \in [N] \setminus I: 
\supp(a_j)\subseteq \cup_{k \in I}\supp(a_k)   \}) \le  \epsilon. \nonumber
\end{align}
In other words, the union of supports of a
randomly and uni\-form\-ly chosen subset of $t$ columns of a $(t,\epsilon)$-disjunct matrix does not contain the support of any
other random column with probability at least $1-\epsilon$. 
\end{definition}

Note that, this definition crucially differs from that of $(t,\epsilon)$-disjunct matrices appeared in \cite{mazumdar2015}. In the definition of
\cite{mazumdar2015},  the probability $P_{R_t}(\{I \in \cP_t(N), \,\,  \forall j \in [N] \setminus I: 
\supp(a_j)\subseteq \cup_{k \in I}\supp(a_k)   \})$ must be upper bounded by $\epsilon$, instead of what we have.

The  next fact follows from the definition of disjunct matrix and the decoding procedure \cite[p.~134]{du1993combinatorial}.
\vspace*{.05in}\begin{proposition}\label{prop:scheme}
A  $(t,\epsilon)$-disjunct matrix defines a group testing scheme that can identify
all items in a random defective configuration of size  $t,$ and with probability at most $\epsilon$ identifies
any randomly chosen item outside of the defective configuration as defective (false-positive).
\end{proposition}
\vspace*{.05in}\begin{remark}
\textcolor{black}{This definition implies that the probability that there is a false positive is bounded above by $\epsilon (N-t)$, 
so as $N$ increases, we need that $\epsilon$ goes to zero at least linearly with $N$.}

For a fixed $N$, unless $\epsilon < \frac{t}{N-t}$, the average number of false positives in the scheme given by a $(t,\epsilon)$-disjunct matrix will be greater than the actual number of defectives. However even in that case, the tests will output a subset of $[N]$ of a vanishingly small proportion that includes all
of the $t$ defective items.

\end{remark}

\subsection{Codes and the Kautz-Singleton construction} 
   \label{sect:KS}
Let $\cQ=\{a_1,\dots,a_q\}$ be a finite set (an alphabet). A code of length $M$ is a subset of the set $\cQ^M$. The minimum Hamming distance between distinct codewords of $\cC$ is called the distance of the code. We use the notation $(M,N,d)$ to refer to a code $\cC$
of length $M$, cardinality $N$ and distance $d$. If $\cQ$ is a finite field and $\cC$ is a linear subspace of the vector space $\ff_q^M$, we
call $\cC$ a linear code. If all the codewords of the code $\cC$ contain exactly $w$
nonzero entries, we call it a constant weight code and use the notation $(M,N,d,w)$ (clearly, constant weight codes cannot be linear).

Kautz and Singleton \cite{kautz1964nonrandom} made two observations. First, they noted that constant weight binary codes give rise to 
disjunct matrices. More precisely, the following is true.
\begin{proposition}\label{prop:disj}
An $(M,N,d,w)$ constant weight binary code $\cC$ provides
a $t$-disjunct matrix, where
$
t = \Big\lfloor\frac{w-1}{w-d/2}\Big\rfloor.
$
\end{proposition}
\vspace*{.1in}\begin{IEEEproof} Write the codewords of $\cC$ as the columns of an $M\times N$ matrix.
The intersection of supports of any two columns has size at most $w -d/2$. Hence if
$w> t(w-d/2)$, the support of any column will not be contained in the union of supports of any $t$ other columns.
\end{IEEEproof}

\vspace*{.1in}This proposition implies that a group testing scheme can be obtained from constant weight codes with large
distance. In \cite{kautz1964nonrandom}, it is observed that such codes can be obtained from non-constant-weight
$q$-ary codes in which every symbol is replaced by its binary indicator vector in the alphabet.
More formally, let $c=(c_1,\dots,c_M)\in \cC$ and let $c_i=a_{j_i},$ where $a_{j_i}\in \cQ, i=1,\dots,M.$ The {\em Kautz-Singleton map} transforms
$c$ to a binary vector of length $qM$ by mapping $c_i, i=1,\dots,M$ to a binary vector of length $q$ that contains $1$ in position $a_{i_j}$ and
zeros elsewhere. We note that the image of this map applied to any code is a binary constant weight code, and that applying this map
to any two $q$-ary vectors $c_1,c_2$ results in a pair of binary vectors with Hamming distance twice the distance $d(c_1,c_2).$
Applying the Kautz-Singleton map to a $q$-ary $(n,N,d)$ code $\cC,$ we obtain a constant weight code with the parameters $(M=qn,N,2d,w=n=M/q).$

\subsection{Distance distribution of codes}\label{sect:dd}
Let us recall some concepts related to the distance distribution of codes. 
More information about them can be found, for instance, in \cite{del73,del73a} or \cite{MS1977}.
Let $\cC\subset \ff_q^n$ be a code. The distance distribution
of $\cC$ is the set of numbers $(A_0,A_1,\dots,A_n),$ where 
   $$
   A_i=\frac1{|\cC|}|\{(x,y)\in\cC^2: d(x,y)=i\}|, \quad i=0,1,\dots,n
   $$ 
   is the average number of ordered pairs of codewords with Hamming distance $i.$ Clearly, the distance of $\cC$ equals the smallest $i\ge 1$ such that $A_i>0$.

Define the {\em dual distance} $d'$ of $\cC$ as follows:
 \begin{equation}\label{eq:dualdeff}
    d'(\cC)=\min \{j\ge 1 : A_j':=\sum_{i=0}^n A_i K_j(i)>0\}, 
   \end{equation}
where $K_j(i)$ is the value of the Krawtchouk polynomial of degree $j$ \cite[p.129]{MS1977} given by
  $$
  K_j(i)=\sum_{l=0}^j (-1)^l \binom il \binom {n-i}{j-l}(q-1)^{i-l}.
  $$
If $\cC$ is a linear code, then $d'$ equals the distance of the dual code $\cC^\bot.$ We also note that $A_0'=1.$

Similar definitions can be given for binary constant weight codes. Let $J_M^w$  
   be the set of all binary vectors (of length $M$)
   with $w$ ones and let $\cC\subset J_M^w$ be a code. We use the notation $\cC(M,N,d,w)$ to refer to a constant weight code of 
   length $M$, cardinality $N,$ distance $d$ and weight $w$. Define the distance distribution of the constant weight code
$\cC$ as follows: 
  \begin{equation}\label{eq:bi}
   b_i=\frac{1}{|\cC|}|\{(x,y)\in \cC^2: w-|\supp(x)\cap \supp(y)|=i\}|
   \end{equation}
   for $ i=0,1,\dots,w$.
The {\em dual distance} $d'$ of the constant weight code $\cC$ is defined as
 $$
  d'(\cC)=\min\Big\{ j\ge 1: b_j':=\frac1{|\cC|}\sum_{i=0}^w b_i Q_j(i)>0\Big\},
 $$
 where $Q_j(i)$ is the value of the Hahn polynomial of degree $j$; see \eqref{eq:H} and \cite[p.545]{MS1977}.

As is well known \cite{del73}, a binary constant weight code with dual distance $d'=r+1$ forms a combinatorial design of strength $r.$

\vspace*{.05in}\begin{definition}\label{def:designs}
An $r$-design (in more detail, an $r$-$(n,w,\lambda)$ design) is a collection $\cC$ of $w$-subsets of an $n$-set $V$, called blocks, such that every $r$ elements of $V$ are contained in the same number $\lambda$ of blocks. An $r$-design is also called a design of strength $r.$
\end{definition}

\vspace*{.05in}  Below we use designs of a given strength to construct almost disjunct matrices. The use of combinatorial designs for constructing disjunct matrices is not new, see, e.g., Sect. 7.4 of \cite{du1993combinatorial}. Special cases of designs have been used to construct nonadaptive group testing matrices for some particular 
parameters \cite[Sec.~11.3]{stinson2004combinatorial}, \cite[Ch.56]{HCD07}. However the conclusion in \cite[p.~146]{du1993combinatorial},  is that disjunct matrices obtained from designs are of little interest because the number of tests $M$ in such constructions is too large compared to the number of subjects $N$. \textcolor{black}{All of the cited works estimate
the parameters of the design that give rise to a disjunct matrix. We take a different approach, using designs formed
by codes to analyze the distance distribution of codes and estimate the probability that a chosen $t$-set
of columns violates the disjunct condition. As a result, we are able to construct almost disjunct matrices with
a small number of tests $M$ compared to $N$.}

\subsection{Sums of independent random variables}
Here we list several classical results about sums of independent random variables and bounds on the moments of such sums.
We begin with an inequality due to Hoeffding.
\begin{theorem}{\cite[Thm.~4]{hoeffding1963probability}}\label{thm:h} Let $C$ be a finite set of numbers. Let $X_i, i=1,\dots, t$ be random samples 
without replacement from $C$ and let $Y_i, i=1,\dots, t$ be random samples with replacement from $C$. If the function $f(x)$
is continuous and convex, then
    \begin{equation}\label{eq:AV}
    \avg f\Big(\sum_{i=1}^t X_i\Big)\le \avg f\Big(\sum_{i=1}^t Y_i\Big).
    \end{equation}
\end{theorem}
We also use the following inequalities for moments of sums of independent random variables.
Let $\{X_i, i=1,\dots,t\}$ be a sequence of independent random variables with zero means.
Then the following {\em {M}arcinkiewicz--{Z}ygmund inequality} holds true \cite{ren2001best}:
  \begin{equation}\label{eq:MZ}
  \avg\Big|\sum_{i=1}^t X_i\Big|^\ell\le C_\ell t^{\ell/2-1}\sum_{i=1}^t \avg|X_i|^\ell, \quad \ell \ge 2,
  \end{equation}
where $C_\ell$ depends only on $\ell$. In particular, one can take $C_l\le (3\sqrt 2)^l l^{l/2}.$
Assume in addition that $E|X_i|^\ell<\infty$ for all $\ell>2.$ Then the following {\em Rosenthal inequality} holds true \cite{Rosenthal70}:
  \begin{align}
  \max\Big\{\sum_{i=1}^t \avg|X_i|^\ell,&\Big(\sum_{i=1}^t\avg|X_i|^2 \Big)^{\ell/2}\Big\} \notag
  \\
 & \le \avg \Big|\sum_{i=1}^t X_i\Big|^\ell\le K_\ell\max\Big\{\sum_{i=1}^t \avg|X_i|^\ell,\Big(\sum_{i=1}^t\avg|X_i|^2\Big)^{\ell/2}\Big\},\label{eq:Ros}
  \end{align}
  where the factor $K_\ell$ does not depend on $t$. Moreover, one can take $K_\ell =({2\ell}/{\log \ell})^\ell$ \cite{Johnson85}.

Observe that inequalities \eqref{eq:MZ}, \eqref{eq:Ros} belong to a large group of Khinchine-type inequalities and
their extensions to martingales. It is possible to further optimize the constant in \eqref{eq:Ros} as well as to establish other
versions of \eqref{eq:MZ} and \eqref{eq:Ros}, see e.g., \cite{Johnson85,Peshkir95}. The choice of the inequality depends on the relation between
the parameters of the group testing scheme, and below we do not attempt to optimize the constants for the large number of possible cases.

\section{Almost disjunct matrices from codes}\label{sect:mc}

\vspace*{.05in}

The main contribution of this paper is a refined analysis of the {\em distance
distribution} of codes that gives rise to almost disjunct matrices (see Def.~\ref{def:adm}). The dual distance
of the code defined in sections \ref{sect:AG} and \ref{sect:CW} plays an important role in the analysis.

Our goal is to design an $M \times N$ matrix $A$ such that 
$t$  randomly chosen columns do not contain the support of any other of its columns. As above, we form the
matrix using the codewords of an $(M,N,d,w)$ constant weight code $\cC$ as the columns.
Let the codewords of $\cC$ be $x_1,x_2,\dots, x_{N}$ and let $d_{ij}:=d(x_i,x_j).$
Having in mind the design of almost disjunct matrices, we begin with the following extension of Prop.~\ref{prop:disj}.
Denote by $P_A(t,N)$ the probability of violating the conditions of Def.~\ref{def:adm}:
$$
  P_A(t,N):=P_{R_t'}(\{I \in \cP_t(N), j \in [N] \setminus I:
\supp(x_j)  \subseteq \cup_{k \in I}\supp(x_k)\}).
$$
It is clear that, any matrix $A$ is $(t, P_A(t,N))$-disjunct.
\begin{proposition}\label{prop:pks} The following estimate holds true:
\begin{gather}
P_A(t,N)\le
P_{R_t'}\Big(\Big\{I \in \cP_t(N), j \in [N] \setminus I: w \le \sum_{k \in I} (w -d_{jk}/2)\Big\}\Big).
\label{eq:pks}
\end{gather}
\end{proposition}
In the next sections we develop new ways of analyzing almost disjunct matrices from
various families of codes, and also connect them with combinatorial designs.
In particular, we examine two different ways of constructing almost disjunct matrices from codes using the above proposition.

\subsection{Almost disjunct matrices from nonbinary codes}\label{sect:AG}
In this section we estimate the probability $P$ in \eqref{eq:pks} for nonbinary linear codes used in the Kautz-Singleton construction.

Consider a $q$-ary $(n,N,d)$ $\cC$. To construct a group testing (almost disjunct) 
matrix we map every symbol of the codeword to a binary vector of $(q-1)$ $0$s and one $1$ in the location that corresponds to the value of the symbol. 
Applying this mapping, we obtain a set of  binary vectors of length $M$ and constant weight $w =n = M/q.$ 
The parameters of the resulting binary constant weight code $\cD$ are $(M=qn,N,2d,w=n=M/q).$

The main result proved in this part is given by the following statement.
\begin{theorem}\label{prop:codes} Let $\cC$ be a  $q$-ary $(n,N)$ code with dual 
distance $d'$ and let $\mathcal M$ be an $M\times N$ matrix constructed from it using the Kautz-Singleton mapping.
 {If $t\le q$, then} 
 $\mathcal M$ is a $(t,\epsilon)$-disjunct matrix with 
\begin{equation}
\epsilon  \le B(\ell,t)\Big(\frac{ e\ell(q-1)}{2n (q-t)^2}\Big)^{\ell/2} \sum_{i=0}^{\ell/2} \Big(\frac{(q-1)\ell}{2ne}\Big)^i, \label{eq:duald1}
\end{equation}
 for any even $\ell<d'$, where $B(\ell,t)=\min\big\{(18\ell t)^{\ell/2}, t^\ell\big\}.$
\end{theorem}
%

\vspace{0.1in}The examples given below show that it is possible to choose specific code families so that estimate 
\eqref{eq:duald1} is 
nontrivial. To prove Theorem \ref{prop:codes} we need several auxiliary statements.

\vspace{0.1in}Choose two codewords from the code $\cC$ randomly and uniformly with replacement, and denote by $Z$ the random variable
whose value is the distance between these codewords.
Clearly, the distribution of $Z$ is given by
   $$
\Pr(Z=i) = \frac{A_i}{N}, \quad i=0, \dots, n.
   $$
 Define $\theta:=(q-1)/q.$

We have the following proposition.
\begin{proposition} \label{prop:moments}
Let $t\le q.$ Then 
\begin{align}
P_A(t,N)\le B(\ell,t)\Big(\frac{q}{n(q-t)}\Big)^\ell\avg(\theta n- Z)^\ell, \label{eq:moment}
\end{align}
where $\ell\ge 2$ is an even integer. 
\end{proposition}
\begin{IEEEproof}
Let $\cC$ be the $q$-ary code defined before the theorem, and let $\cD$ be the code obtained from $\cC$ by
applying the Kautz-Singleton mapping. Given two codewords $x_j,x_k\in \cC$ let $\delta_{jk}=d(x_j,x_k)$ and let
$d_{jk}=2\delta_{jk}$ be the distance between their images in $\cD.$ 
We will estimate from above the right-hand side in \eqref{eq:pks}. \textcolor{black}{With the current notation, the condition
in \eqref{eq:pks} becomes $\sum_{k\in I}\delta_{jk}\le n(t-1).$ Using the assumption $t\le q$ yields $1-\frac{t}{q} \ge 0,$ and we can relax the inequality in \eqref{eq:pks} to the following estimate:}
  \begin{align*}
  P_A(t,N)&\le  P_{R_t'}\Big(\Big\{I \in \cP_t(N), j \in [N] \setminus I:  \sum_{k \in I} (\theta n -\delta_{jk}) \ge n \Big(1-\frac t q\Big)
   \Big\}\Big)\\
   &\le P_{R_t'}\Big(\Big\{I \in \cP_t(N), j \in [N] \setminus I:  \Big(\sum_{k \in I} (\theta n -\delta_{jk})\Big)^\ell \ge 
 n^\ell \Big(1-\frac t q\Big)^\ell\Big\}\Big)\\
      &\le  \Big(n-\frac {nt} q\Big)^{-\ell}\avg\Big( \sum_{k \in I} (\theta n -\delta_{jk})\Big)^\ell.
  \end{align*}
Here the second line follows because $\ell$ is even and on the third line we use the Markov inequality.
Observe that $\delta_{jk}$ are random variables that denote the distance between two codewords $x_j$ and $x_k, 1 \le k \le t$ chosen 
randomly from without replacement.  Let $\mu_{jk}$ denote the random variable corresponding to $\delta_{jk}$ when the codewords are chosen with replacement, and note
 that $\mu_{jk}$ and $\mu_{j',k'}$ are independent whenever $j\ne j'$ or $k\ne k'$.
 Using Theorem \ref{thm:h}, we have
   \begin{align}\label{eq:b}
   P_A(t,N)\le \Big(n-\frac {nt} q\Big)^{-\ell}{\avg\Big( \sum_{k \in I} (\theta n -\mu_{jk})\Big)^\ell}.
   \end{align}
Let us estimate the numerator in \eqref{eq:b} using the Minkowski inequality (the triangle inequality for the $\ell$-norm), 
again keeping
in mind that $\ell$ is even. 
We obtain
  \begin{align*}
P_A(t,N)& \le \frac{\Big(  \sum_{k \in I} \Big(\avg (\theta n -\mu_{jk})^\ell\Big)^{1/\ell} \Big)^\ell}{n^\ell(1-\frac{t}{q})^\ell}= \frac{t^\ell \avg(\theta n- Z)^\ell}{n^\ell(1-\frac{t}{q})^\ell},
  \end{align*}
Now let us estimate the right-hand side of \eqref{eq:b} relying on the {M}arcinkiewicz--{Z}ygmund inequality \eqref{eq:MZ}.
\remove{ It cannot be immediately applied
because, though the random variables in \eqref{eq:b} have zero means, they are dependent. 
However, since $\ell$ is even, the function $(\cdot)^\ell$ is convex, so we can bound \eqref{eq:b} above by a sum of independent random variables 
using \eqref{eq:AV}, and then use inequality \eqref{eq:MZ}. }
 We obtain
   \begin{align*}
   P_A(t,N) \le \frac{(3\sqrt{2})^\ell \ell^{\ell/2}t^{\ell/2-1}\sum_{k \in I} \avg (\theta n -\mu_{jk})^\ell}{n^\ell(1-\frac{t}{q})^\ell}= \frac{(18\ell t)^{\ell/2} \avg(\theta n- Z)^\ell}{n^\ell(1-\frac{t}{q})^\ell}.
   \end{align*}
This completes the proof.
\end{IEEEproof}

Our next step will be to estimate the moments in \eqref{eq:moment}.
\begin{lemma}
Let $\cC$ be a  code over $\ff_q$ of length $n$ and size $N$ with dual distance
$d'.$
For any $\ell<d'$,
\begin{align}
\frac{1}{N}&\sum_{j=0}^n  (j-\theta n)^\ell A_j = 
\sum_{j=0}^n (j-\theta n)^\ell \binom{n}{j}\theta^j (1-\theta)^{n-j}.
\label{eq:pless}
\end{align}\label{lem:pless}
\end{lemma}
\begin{IEEEproof}
Let $\cC\subset \ff_q^n$ be a $q$-ary code with distance distribution
$A_0,A_1,\dots,A_n$ and let $A'_i,i=0,1,\dots,n$ be the dual distance distribution defined in \eqref{eq:dualdeff}.
The following {\em Pless power moment identities} hold true \cite{pless1963power},\cite[p.131]{MS1977}:
  \begin{equation}\label{eq:PPMI}
  \sum\limits_{j=0}^n j^r A_j  = \sum\limits_{j=0}^n (-1)^j A_j'\Big(\sum\limits_{\nu =0}^r\nu!S(r,\nu)q^{k-\nu}(q-1)^{\nu-j}\binom{n-j}{n-\nu}\Big)
  \end{equation}
where $S(r,\nu)$ are the Stirling number of second kind,
$$
S(r,\nu) = \frac{1}{\nu!}\sum_{i=0}^\nu (-1)^{\nu-i}\binom{\nu}{i}i^r,
$$
when $\nu \le r$ and $S(r,\nu) =0$ otherwise.
Taking $r<d',$ we find that on the right-hand side of \eqref{eq:PPMI} the sum on $\nu$ is zero unless $j<d',$ but then
$A_j'=0$ except for $A_0'=1.$ 
Using this in \eqref{eq:PPMI}, we obtain
$$
\sum_{j=0}^n j^r A_j  = \sum_{\nu =0}^r\nu!S(r,\nu)q^{k-\nu}(q-1)^{\nu}\binom{n}{n-\nu}.
$$
Substituting the definition of $S(r,\nu)$ and simplifying, 
we find that 
$$
\sum_{j=0}^n j^r A_j  = \sum_{j =0}^n j^r\binom{n}{j}\theta^j (1-\theta)^{n-j}.
$$
The lemma follows immediately.
\end{IEEEproof}

Another proof would be to use the general result for symmetric association schemes \eqref{eq:f} established below together
with the properties of the Hamming association scheme. We believe that readers familiar with association schemes will have no difficulty filling in
the details.

Let us bound above the right hand side of \eqref{eq:pless}.

\begin{lemma}
Let $1/2<p<1$. Then,  
\begin{align*}
\mu_n(2r) & := \sum_{j=0}^n \biggl(\frac{j-np}{\sqrt{p(1-p)}}\biggr)^{2r} \binom{n}{j}p^j (1-p)^{n-j}
< (ner)^r \sum_{i=0}^r \Big(\frac{pr}{(1-p)ne}\Big)^i.
\end{align*}
\label{lem:approx}
\end{lemma}
\begin{remark}\label{remark:2r}
When $\frac{p}{1-p} \le \frac{n}r$, the expression on the right hand side above  can be further simplified.
Namely, $\frac{pr}{ne(1-p)} \le e^{-1}$, and we have,
$$
\mu_n(2r) \le \frac{1}{1-e^{-1}}(ner)^r.
$$
\end{remark}
\begin{IEEEproof}[Proof of Theorem~\ref{prop:codes}]
Using Lemma  \ref{lem:pless} and  Lemma \ref{lem:approx} with $p = \theta$, we immediately obtain, for any even $\ell < d',$
\begin{equation*}
\avg(Z - \theta n)^\ell \le 
\Big(\frac {ne\ell\theta}{2q}\Big)^{\ell/2}\sum_{i=0}^{\ell/2} \Big(\frac{(q-1)\ell/2}{ne}\Big)^i .
\end{equation*}
Substituting this estimate in \eqref{eq:moment} and rearranging, we obtain the bound \eqref{eq:duald1}.
\end{IEEEproof}

\subsection{Examples}\label{sec:examples}
Let us examine a few specific choices of the outer $q$-ary codes. 
In each case our goal is to choose the parameters so that the quantity in \eqref{eq:duald1} is small, and to
examine the parameters 
of the resulting almost disjunct matrices and  group testing schemes.

\vspace{0.1in} {\em 1) Reed-Solomon codes:} 
In this case, $n =q-1$, $\ell\approx\log_q N$. \textcolor{black}{If $t> 18\log N$, then the minimum in $B(\ell,t)$ is attained for the
first term, otherwise for the second one. For instance, consider the first option.}
From \eqref{eq:duald1} we obtain
\begin{align}\label{eq:Ntq}
P_A(t,N)
&< \Big(\frac{9e\ell^2 t}{(q-t)^2}\Big)^{\ell/2} \Big(\frac{\ell}{2e}\Big)^{\ell/2+1}
  \approx\frac{\ell}{2e}\Big(\frac{2.13\ell^{3/2}\sqrt{t}}{q-t}\Big)^\ell.
\end{align}
Thus the probability $P$ is small if we take $q>2.13\ell^{3/2}\sqrt t+t$.
 Note that for Reed-Solomon codes we have $M=q(q-1).$ Overall we obtain
  $
  M=O\big(\max\{t^2 , t(\log_q N)^3\}\big).
  $

\vspace{0.1in}
{\em 2) Algebraic-geometric codes:}
Let us consider two examples that rely on codes on algebraic curves.
\paragraph{Hermitian codes}
Let $q_0$ be a power of a prime and let $r$ be an integer such that $$q_0^2-q_0-2\le r\le q^3.$$
There exists a family of linear $q$-ary codes, $q=q_0^2,$ constructed on Hermitian curves \cite[Sec.8.3]{sti93}. The parameters 
of the codes are as follows: 
  \begin{gather*}
  \text{length } n=q_0^3, \quad\text{cardinality } N=q_0^{2(r+1-q(q-1)/2)}\\
  \text{dual distance }d'\ge r+q_0+2-q_0^2.
  \end{gather*}
In particular, choosing $r=q_0^2,$ we obtain $d'\ge q_0+2$ and $N\approx q_0^{2q_0^2}.$
This suffices to ensure that the quantity in \eqref{eq:duald1} is small, i.e.,
the matrix formed by using Hermitian codes in the Kautz-Singleton construction
is almost disjunct. Assuming that the number of defectives $t<q_0^2,$ we obtain $M=q_0^5$ for the
number of tests. 

\vspace*{.1in}
\paragraph{Suzuki codes}
Similar results are obtained if we take codes constructed on Suzuki curves \cite{Hansen90}. Namely, let $q=2q_0^2, q_0=2^m$ and let $r$ be an integer
such that $2q_0(q-1)-2<r<q^2.$ There exists a family of linear $q$-ary codes with the parameters
  \begin{gather*}
  \text{length } n=q^2, \quad\text{cardinality } N=q^{r+1-q_0(q-1)}\\
  \text{dual distance }d'\ge r-2(q_0(q-1)-1).
  \end{gather*}
Choosing $r=2q_0q,$ we obtain $d'\ge 2q_0+2,$ so we can take $\ell=(1/4)n^{1/4}.$ Substituting this in \eqref{eq:duald1}, we see that $P \sim n^{-\ell/4}$
for any  $t\leq q/2,$ which allows us to choose the number of defectives $t=O(n^{1/2})=O(q).$ 
With this choice of the parameters, testing matrices obtained
from Suzuki codes using the Kautz-Singleton construction are guaranteed to have the almost disjunct property.
With the above choice of $r$ we have $t=q/2$,
   $N=q^{2q_0^3+q_0+1}$ and
   $M=nq=q^3.
   $

\subsection{Almost disjunct matrices from constant weight codes}\label{sect:CW}
In this part we study properties of matrices constructed from constant weight codes with a known value of the dual distance $d',$
i.e., combinatorial designs of strength $d'-1$. 

The scheme of the proof is similar to the previous section. Beginning with 
Prop.~\ref{prop:pks} we will estimate the probability of a false positive using moments of the distance distribution of constant weight codes. First let us show that if $r<d',$ then the $r$th central moment of distance distribution equals the $r$th moment of
the distance distribution of the sphere of weight $w$.

\vspace*{.05in}
\begin{theorem} \label{thm:moment} Let $\cC$ be a constant weight code of weight $w$, length $M$, distance distribution $\{b_i, i =0, \dots, w\}$ and
dual distance $d'$.  Let $X$ be a hypergeometric random variable 
with pmf and moments
     \begin{equation}\label{eq:pmf}
     f_X(i)=\frac{\binom wi\binom{M-w}{w-i}}{\binom Mw}, \quad i=0,1,\dots, w; \quad \avg X=\frac{w^2}M, \;\Var(X)=
     \frac{w^2(M-w)^2}{M^2(M-1)}.
     \end{equation}
 As long as $r<d',$
    \begin{equation}\label{eq:hyp}
   \sum_{i=0}^w \Big( \frac{w(M-w)}M- i\Big)^r b_i=|\cC|\;\avg((X-\avg X)^r).
   \end{equation}
\end{theorem} 
\begin{IEEEproof}
See the appendix.
\end{IEEEproof}

\vspace*{.05in}\begin{remark}
Note that $\vartheta:=w(M-w)/M$ is the average pairwise distance in the set of all binary vectors of weight $w,$ so 
equality \eqref{eq:hyp} gives 
the moments of the distance distribution about the mean. In this sense it is analogous to the corresponding result for the Hamming space \eqref{eq:pless}.
Irrespective of the value of $r$, the left-hand side of \eqref{eq:hyp} is always greater than or equal to the
right-hand side (this can be seen from \eqref{eq:f} and \eqref{eq:S} in the Appendix). This result was first proved in \cite{sid75} using analytic methods, and is known as the Sidelnikov inequality. 
A similar inequality for general symmetric association schemes is proved in \cite[p.55]{tar82} using a combinatorial approach 
which we adopt in our proof of Theorem \ref{thm:moment}. 
\end{remark}

\vspace*{.05in}
Let $Z:=(1/2)d(x,y)$, where $d$ is the Hamming distance, be the random variable defined by two random vectors chosen from a constant
weight code (with replacement). We have $\Pr(Z=i)=b_i/|\cC|.$ Moreover, if $d'\ge 3$ then Theorem \ref{thm:moment} implies
that the central moments of $Z$ up to order $d'-1$ are the same that the central moments of $X,$ so
  \begin{equation}\label{eq:EV}
  \avg Z=\vartheta, \quad \Var(Z)=\vartheta^2/(M-1).
  \end{equation}

\vspace*{.05in} The main result of this section is given in the next theorem.
\begin{theorem}\label{prop:prop10}
Let $\cC$ be an $(M,N,d,w)$ constant weight code with dual distance $d'$ and let $w<M/2$. 
Let $t$ be the maximum number of defective items and suppose that $t<M/w$. 
For any even $\ell<d'$ the group testing scheme constructed from $\cC$
is $(t,\epsilon)$-disjunct for
\begin{align}\label{eq:const_wt}
\epsilon&\le {B}(\ell,t) 
\Big(\frac{e\ell(M-w)}{2(M-tw)^2}\Big)^{\ell/2}\sum_{i=0}^{\ell/2} \Big(\frac{(M-w)\ell}{2ew^2}\Big)^{i},
\end{align}
where $B(\ell,t)=\min\{(18\ell t)^{\ell/2}, t^\ell \}.$ In addition, if $M \ge \max\{ 4w^2t/\ell^2, w +2ew^2/\ell\}$ 
then the matrix is $(t,\epsilon)$-disjunct with 
   \begin{align}\label{eq:PRos}
   \epsilon\le t\Big(\frac{2\ell^2{(M-w)} }{(\log \ell) w (M-tw)}\Big)^\ell.
\end{align}
In the case of $\ell=2$ we can use the following bound instead of the above,
  \begin{equation}\label{eq:l=2}
  \epsilon<\frac t{M-1}\frac{(M-w)^2}{(M-wt)^2}.
  \end{equation}
\end{theorem}

{\em Remarks:} 
We comment on the conditions for the bounds \eqref{eq:const_wt}-\eqref{eq:PRos} to be nontrivial in Corollary
\ref{cor:<1} below. As for \eqref{eq:l=2}, it gives a good bound for large $M$ and small or slowly growing $w$ and $t$.

\vspace*{.1in}
\begin{IEEEproof}
For all $ k \in I$ and an random index $j \in [N]\setminus I$,  let $\xi_{jk}$ be
the random value of $d_{jk}/2 = d(x_j,x_k)$ when the vectors $x_j,x_k , k\in I$ are chosen from $\cC$ randomly and uniformly without replacement
and let $\eta_{jk}$ denote the same quantity when the vectors are chosen with replacement. The variables $\eta_{jk}, \eta_{jk'}$ 
are independent whenever $k\ne k'$, and each of them is stochastically equivalent to the random variable $Z$
defined above.
Proceeding similarly to the proof of Proposition \ref{prop:moments}, we obtain
\begin{align}
P_A(t,N)&\le P_{R_t'}\Big(\Big\{I \in \cP_t(N), j \in [N] \setminus I: w \le \sum_{k \in I} (w -\xi_{jk})\Big\}\Big)\nonumber\\
& = P_{R_t'}\Big(\Big\{I \in \cP_t(N), j \in [N] \setminus I: \sum_{k \in I} 
    (\vartheta -\xi_{jk}) \ge w\Big(1-\frac{tw}{M}\Big)\Big\}\Big)\nonumber\\
& \le \Big(w-\frac{tw^2}{M}\Big)^{-\ell}\avg\Big(\sum_{k \in I} (\vartheta -\xi_{jk})\Big)^\ell\nonumber\\
&\le \Big(w-\frac{tw^2}{M}\Big)^{-\ell}\avg\Big(\sum_{k \in I} (\vartheta -\eta_{jk})\Big)^\ell,\label{eq:l2}
\end{align}
where in the last step we used  \eqref{eq:AV} and where $\ell, 2\le \ell<d'$ is an even number. 

Now let us use the Minkowski inequality to obtain
\begin{align}
P_A(t,N)& \le \Big(w-\frac{tw^2}{M}\Big)^{-\ell}\Big(\sum_{k \in I} (\avg (\vartheta -\eta_{jk})^\ell)^{1/\ell}\Big)^\ell \nonumber\\
& = \Big(w-\frac{tw^2}{M}\Big)^{-\ell}{t^\ell \avg (\vartheta -Z)^\ell}\nonumber\\
& \stackrel{\eqref{eq:hyp}}=   \Big(w-\frac{tw^2}{M}\Big)^{-\ell}{t^\ell\avg((X-\avg X)^\ell) },\label{eq:const_bound1}
\end{align}
where $X$ is the hypergeometric random variable defined by \eqref{eq:pmf}.

As before, we can also estimate the right-hand side of \eqref{eq:l2} relying on the Marzinkiewicz-Zygmund inequality
\eqref{eq:MZ}. We obtain
\begin{align*}
P_A(t,N) &\le \Big(w-\frac{tw^2}{M}\Big)^{-\ell}{\avg\Big(\sum_{k \in I} (\vartheta -\eta_{jk})\Big)^\ell}\\
& \le \Big(w-\frac{tw^2}{M}\Big)^{-\ell}{(3\sqrt{2})^\ell \ell^{\ell/2}t^{\ell/2-1}\sum_{k \in I} \avg(\vartheta -\eta_{jk})^\ell}\\
& \le \Big(w-\frac{tw^2}{M}\Big)^{-\ell}{(18\ell t)^{\ell/2} \avg(\vartheta -X)^\ell}.
\end{align*}
Taking this estimate together with \eqref{eq:const_bound1} we obtain the bound
\begin{align}
P_A(t,N) \le \min\{(18\ell t)^{\ell/2}, t^\ell \} \frac{ \avg(\vartheta -X)^\ell}{w^\ell(1-\frac{tw}{M})^\ell}.\label{eq:const_bound_}
\end{align}

It is evident that $X =\sum_{i=1}^w X_i$, where $X_1, \dots,X_w$ denote values of random samples drawn without 
replacement from a population of $M$ values with $w$ ones and $M-w$ zeros. Now using Theorem \ref{thm:h}
we claim that
  \begin{align}
  \avg((X-\avg X)^\ell) \le \avg((Y-\avg Y)^\ell),
  \end{align}
   where $Y \equiv \sum_{i=1}^w Y_i$, and $Y_1, \dots,Y_w$ denote samples 
from the same population with replacement. This makes the $Y_i$'s into independent Bernoulli($\frac wM$) random variables, and
\begin{align} 
\avg((Y-\avg Y)^\ell) &= \sum_{j=0}^w \Big(j -\frac{w^2}{M}\Big)^\ell\binom{w}{j}\Big(\frac wM\Big)^j\Big(1-\frac wM\Big)^{w-j}\nonumber\\
& = \sum_{j=0}^w \Big(w\Big(1-\frac{w}{M}\Big) -j\Big) ^\ell\binom{w}{j}\Big(\frac{w}{M}\Big)^{w-j}\Big(1-\frac{w}{M}\Big)^{j}\nonumber\\
& < \Big(\frac{w^2e\ell}{2M}\Big(1-\frac{w}{M}\Big)\Big)^{\ell/2}\sum_{i=0}^{\ell/2} \Big(\frac{(M-w)\ell}{2ew^2}\Big)^{i},\label{eq:?}
\end{align}
where in the last step we have used Lemma~\ref{lem:approx} with $p =1-w/M$. 
Using \eqref{eq:?} in \eqref{eq:const_bound_} and rearranging,
we obtain for the probability of the false positive the estimate \eqref{eq:const_wt}.

Before proving \eqref{eq:PRos}, consider the case $\ell=2.$ Here the calculation is simpler because we can directly
substitute the value of the variance of $Z.$ 
Indeed, by independence
   \begin{align}
 \avg\Big(\sum_{k \in I} (\vartheta -\eta_{jk})\Big)^2
 =\sum_{k\in I}\Var(\eta_{jk})
   =\frac {t\vartheta^2}{M-1}. \label{eq:leq2}
  \end{align}
 Using this in \eqref{eq:l2} and simplifying, we obtain \eqref{eq:l=2}.

Now let us prove \eqref{eq:PRos}, estimating the moment of the sum of random variables using
Rosenthal's inequality \eqref{eq:Ros}. We have
  \begin{equation}\label{eq:Ros1}
  \avg\Big(\sum_{k\in I}(\vartheta-\eta_{jk})\Big)^\ell\le K_\ell\max\Big\{\sum_{k\in I}\avg(\vartheta-\eta_{jk})^\ell,
    \Big(\sum_{k\in I}\avg(\vartheta-\eta_{jk})^2\Big)^{\ell/2}\Big\}
    \end{equation}
To estimate the maximum, we will take an upper bound on the first term given by \eqref{eq:?} and show that it is greater than the second
term. 
Assume that  $M \ge w +2ew^2/\ell$, 
then the largest term under the
sum in \eqref{eq:?} is the last one, and we obtain 
  $$
  \avg((Y-\avg Y)^\ell)\le \frac{\ell}{2}\Big(\frac{\ell(M-w)}{2M}\Big)^\ell,
  $$
and therefore,
  \begin{equation}\label{eq:ubb}
  \sum_{k \in I}\avg (\vartheta -\eta_{jk})^\ell\le \frac{t\ell}{2}\Big(\frac{\ell(M-w)}{2M}\Big)^\ell.
  \end{equation}
The value of the second term on the right-hand side of \eqref{eq:Ros1} is found directly from \eqref{eq:leq2} and equals
   \begin{equation}\label{eq:2}
   \Big(\sum_{k \in I} \Var(\eta_{jk})\Big)^{\ell/2}= \Big(\frac{t\vartheta^2}{{M-1}}\Big)^{\ell/2}.
   \end{equation}
As is easily checked, the right-hand side of \eqref{eq:ubb}  is greater than 
 \eqref{eq:2} if $M\ge 4w^2 t/\ell^2.$ Therefore the right-hand side of \eqref{eq:ubb} also provides an upper bound on the maximum in \eqref{eq:Ros}. Substituting it in \eqref{eq:l2}, we obtain the claimed bound \eqref{eq:PRos}.
 \end{IEEEproof}

\vspace*{.05in} In the next corollary we state the conditions for the bounds of Theorem \ref{prop:prop10} 
to guarantee that we obtain almost disjunct matrices. We focus on the estimate \eqref{eq:PRos},
but a similar claim can be also made with respect to the bound \eqref{eq:const_wt}. 
 \vspace*{.05in} \begin{corollary} \label{cor:<1} Suppose that $w> 2\ell^2/\log\ell$ and $M >\max\{4w^2t/\ell^2,w +2ew^2/\ell,wt\log \ell\}.$
Then the codewords of a binary constant weight code of length $M$, cardinality $N$, weight $w$ and dual distance $d'>\ell$
form an $(t,\epsilon)$-disjunct matrix with 
$\epsilon$ approaching zero exponentially with the increase
of $\ell$. 
\end{corollary}
\vspace*{.05in} \begin{IEEEproof}
Under the stated assumptions the probability of a false positive can be bounded above by \eqref{eq:PRos}, and the term $(\cdot)^\ell$
in that expression approaches zero with increasing $\ell.$
\end{IEEEproof}

\vspace*{.05in}
It is possible to consolidate the restrictions in $M$ in this corollary by making further assumptions on the relation of $t$ and $w$,
but we prefer to leave this statement in the most general form that arises from our estimation method.
This corollary also implies that the number of tests $M$ that suffices to for the matrix to be almost disjunct behaves as
$M=O(\max(t \ell^2/\log^2\ell,\ell^3/\log^2\ell, t\ell^2)) = O(\ell^2\max(\ell/\log^2\ell, t)).$ 
Hence, we have the following corollary. 
\vspace*{.05in}
\begin{corollary}\label{cor:last}
A constant weight code of length $M$ and size $N$ with dual distance greater than $\ell$ provides a $(t, \exp(-\ell))$ disjunct matrix with $t = \Omega(M/\ell^2)$ 
\footnote{Note that due to the lower bound in \eqref{eq:Ros} the order relations obtained here cannot be improved within
the frame of the method considered.}.
\end{corollary}
\vspace*{.05in}

The dual distance of a code (or strength of a design) of course depends on its length and size. 
Note in particular that if we assume $\ell = \Omega(\log N)$, then  the probability that there exist false positives can
be made to approach $0$ (using an union bound on the $N-t$ elements) with total number of tests being $M = O(t \log^2 N)$.

\vspace*{.05in}
 It remains to use results about the existence of $\ell$-designs. Major progress in the existence problem has been
achieved in recent years. 
For instance, due to the result of \cite[Theorem 1.3]{kuperberg2013probabilistic} it is known that there exist $\ell$-designs with
$$
\log N \le c \ell \log (cM/\ell),
$$
for some constant $c>0$, which supports the assumption of $\ell = \Omega(\log N)$ made above. 
We note that this result is existential and not an explicit construction. Substituting it into
Corollary \ref{cor:last}, we obtain the following result.
\vspace*{.05in}\begin{corollary}\label{cor:last1}
There exist nonadaptive group testing schemes constructed from combinatorial designs with $O(t \log^2N /\log^2 t)$ tests that can identify all items in a random defective configuration of size  $t$  with probability of false positive for an element at most $\exp(-\log N/\log t).$ 
\end{corollary}


\vspace*{.05in}Another breakthrough result is due to \cite{Keevash14}.
For a design with the parameters $\ell,M,w,\lambda$ to exist it is necessary that
$\binom{w-i}{\ell-i}$ divides $\binom{M-i}{\ell-i}$ for all $i=0,1,\dots,\ell.$ It has been shown in \cite{Keevash14}
that these conditions are also sufficient, i.e., combinatorial designs exist whenever their parameters satisfy the natural
divisibility constraints.
This advance provides a large supply of objects for the construction of almost disjunct matrices following
the analysis in this section.

\section{Outlook}\label{sect:outlook}
\textcolor{black}{We introduced a general method of constructing  explicit almost disjunct matrices from nonbinary and constant weight codes.
The advantage of the introduced approach is related to its universal nature. While there are limitations
on the applicability of the bounds obtained, in many cases they guarantee existence of almost disjunct matrices.
Moreover, under certain assumptions stated in the paper, the constructed matrices rely on a number of tests of the order $O(t \cdot {\rm polylog} N)$, matching
the best known results. Experimentation shows that some of the simple constructions from constant weight codes behave very well
in identifying random defectives. In particular, in \cite{UMB16} we have conducted experiments with testing matrices constructed from constant weight subcodes of binary BCH codes. Simulations showed that the matrices constructed from all the 
vectors of fixed weight of binary BCH codes of length $M=63$ and $127$ and weight $w=3,5,7$ perform consistently better than random binary matrices in terms of the number of false positives. Another observation in \cite{UMB16} concerns the estimates of the probability of false positives $P_A(t,N)$ given by Theorem \ref{prop:prop10}. Namely, while for the described range of the parameters, 
Eqns.~\eqref{eq:PRos} and \eqref{eq:l=2} give nontrivial estimates of $P_A(t,N)$, in particular, better than those implied by \cite{mazumdar2015}, experimental results give much lower rates of false positives. This shows that the methods introduced here fail to account for the actual performance of some constant weight code matrices, leaving the derivation of better bounds as an open problem.}

\section*{Appendix: Proof of Lemma \ref{lem:approx}}
\begin{IEEEproof}
Let us define i.i.d. random variables $X_i$, $i =1,\dots,n$ in the following way:
\begin{align*}
X_i = \begin{cases}\sqrt{(1-p)/p} & \text{ with probability } p\\
-\sqrt{p/(1-p)} & \text{ with probability } 1-p.
\end{cases}
\end{align*}
Note that $\avg X_i =0$ and $\avg X_i^2 =1.$
Let 
$
X= \sum_{i=1}^n X_i.
$
Clearly, with probability $\binom{n}{i}p^i(1-p)^{n-i}$,
$$
X = i\sqrt{(1-p)/p}-(n-i)\sqrt{p/(1-p)} = \frac{i-np}{\sqrt{p(1-p)}}.
$$
Hence,
$$
\mu_n(2r) = \avg X^{2r}.
$$
Let us estimate the right-hand side. We have
$$
\avg X^{2r} = \sum_{i_1,\dots,i_{2r}} \avg X_{i_1}\dots X_{i_{2r}}.
$$
We use the fact that the variables $X_i$ are independent.
If at least one of the $X_{i_j}$s appears only once then the corresponding monomial is zero. 
So it may be assumed that each index appears at least twice in the expectations that contribute to the sum. In particular, there are
at most $r$ distinct $X_{i_j}$s that can appear. Suppose that $r-t$ such terms 
appear. Let $N_t$ be the number of ways one can assign integers $i_1, \dots, i_{2r} \in 
\{1,\dots,n\}$ such that each $i_j$ appears, at least twice and exactly $r-t$ integers
appear.
Using the fact that $X_{i_j}$s have unit variance and $|X_{i_j}|
\le \sqrt{p/(1-p)}$, we have,
$$
\avg X^{2r} \le \sum_{t=0}^r N_t (\sqrt{p/(1-p)})^{2t} = \sum_{t=0}^r N_t \Big(\frac{p}{1-p}\Big)^t.
$$
A crude bound on $N_t$ gives, 
$$
N_t \le \binom{n}{r-t} (r-t)^{2r}\le \Big(\frac{ne}{r-t}\Big)^{r-t}(r-t)^{2r}.
$$ 
Hence,
\begin{align*}
\avg X^{2r} &\le \sum_{t=0}^r (ne)^{r-t} (r-t)^{r+t}\Big(\frac{p}{1-p}\Big)^t \\
&\le (ner)^r \sum_{t=0}^r \Big(\frac{pr}{ne(1-p)}\Big)^t.
\end{align*}
\end{IEEEproof}

\section*{Appendix: Proof of Theorem \ref{thm:moment}}
We prove Theorem  \ref{thm:moment}, assuming we are given a $(n,N,d,w)$ constant weight code.

1. (Johnson association scheme) We will use some simple properties of the Johnson association scheme \cite{del73,ban84}. 
Recall that $J_n^w$ denotes the set of all binary vectors of length $n$ and Hamming weight $w\le n/2.$ Let $\cC\subset J_n^w$ be a code
whose distance distribution $(b_0=1,b_1,\dots,b_w)$ is defined in \eqref{eq:bi} above.  Define the Hahn polynomial
of degree $k=0,1,\dots,w$ by its values as follows:
     \begin{equation}\label{eq:H}
     Q_k(i)=\frac{\mu_k}{v_i} E_i(k), 
  \end{equation}
  where $E_i(x), i=0,1,\dots,w$ is the Eberlein polynomial, and 
\begin{equation}\label{eq:v}
   v_i=\binom wi \binom {n-w}i, \quad \mu_i=\binom ni-\binom {n}{i-1}.
   \end{equation}
are the valencies and the multiplicities of the scheme $J_n^w.$ The polynomials $(Q_k, k=0,1,\dots,w)$ form an orthogonal system on the set $\{0,1,\dots,w\}.$ The explicit expression for $Q_k(i)$ is well known \cite[p.48]{del73},
\cite[pp.~217-220]{ban84}. Below we need only the expression for the constant $Q_0\equiv 1$ and the linear polynomial, given by
      \begin{equation}\label{eq:Q1}
    Q_1(i)=(n-1)\Big(1-\frac {ni}{w(n-w)}\Big).
    \end{equation}
Define the {\em dual distance distribution} of $\cC$ by 
  $$
  b_j'=\frac1{|\cC|}\sum_{i=0}^w b_i Q_j(i), \quad j=0,1,\dots,w 
  $$ 
and notice that $b_0'=1.$   
By {\em Delsarte's inequalities} \cite[Thm.~3.3]{del73} the coefficients $b_j'$ are nonnegative. 
The Hahn polynomials satisfy the relation
  \begin{equation}  \label{eq:Krein}
  Q_i(l)Q_j(l)=\sum_{k=0}^m q_{ij}^k Q_k(l)
  \end{equation} 
where the numbers $q_{ij}^k$ are called the Krein parameters of the scheme. Importantly, we have
   $
   q_{ij}^k\ge 0
   $ 
\cite[Lemma 2.4]{del73}. 
In fact, the matrices $(E_i(k))$ and $(Q_k(i))$, $i,k=0,1,\dots,w$ form the first and the second eigenvalue matrices of the scheme 
$J_n^w.$ This implies the following relations:
   \begin{gather}
       \sum_{k=0}^w E_k(j) Q_i(k)=\binom nw\delta_{ij}\label{eq:orth}\\
       E_k(0)=v_k \label{eq:val}
       \end{gather}
(for the proofs see Thm.~3.5 and Prop.~3.4 of \cite{ban84}).

\vspace*{.1in}\noindent 2. The calculation below is in part inspired by \cite{tar82}. On account of \eqref{eq:Krein} we can write for all $i=1,\dots, w$

    $$
    Q_1(i)^2=\sum_{k=0}^2 q_{11}^k Q_k(i)
    $$
    and generally
    \begin{equation}\label{eq:r}
    Q_1(i)^r=\sum_{k=0}^r \beta_k(r)Q_k(i), \quad (r\ge 2)
    \end{equation}
where $\beta_k(r)\ge 0$ are some nonnegative coefficients that can be easily computed by orthogonality.
Now consider 
    \begin{align}
    \sum_{i=0}^w Q_1(i)^r b_i&=\sum_{i=0}^w \sum_{k=0}^r \beta_k(r)Q_k(i) b_i \nonumber\\
    &=\sum_{k=0}^r \beta_k(r)\sum_{i=0}^w Q_k(i) b_i \nonumber\\
    &=|\cC|\sum_{k=0}^r \beta_k(r) b_k' \nonumber\\
    &=|\cC|\Big(\beta_0(r)+\sum_{k=d'}^r \beta_k(r) b_k'\Big).\label{eq:f}
  \end{align}
Let us compute $\beta_0(r)$. We have
  \begin{align*}
  \sum_{i=0}^w v_i Q_1(i)^r&\stackrel{\eqref{eq:r}}=\sum_{k=0}^r\beta_k(r) \sum_{i=0}^w v_i Q_k(i)\\
    &\stackrel{\eqref{eq:val}}=\sum_{k=0}^r\beta_k(r) \sum_{i=0}^w E_i(0) Q_k(i)\\
    &\stackrel{\eqref{eq:orth}}=\beta_0(r)\binom nw.
  \end{align*}
  Thus, using \eqref{eq:v} we obtain
    $$
    \beta_0(r)=\frac1{\binom nw} \sum_{i=0}^w \binom wi \binom {n-w}iQ_1(i)^r.  
    $$
Let $0\le r\le d'-1,$ then from \eqref{eq:f} and \eqref{eq:Q1} we have
   \begin{equation}\label{eq:S}
   \sum_{i=0}^w \Big( 1- \frac{ni}{w(n-w)}\Big)^r b_i = \frac{|\cC|}{\binom nw} \sum_{i=0}^w \binom wi\binom{n-w} i \Big( 1- \frac{ni}{w(n-w)}\Big)^r
   \end{equation}
    
 \vspace*{.1in}  Equality \eqref{eq:hyp} follows upon rewriting the right-hand side of \eqref{eq:S} as follows:
      \begin{align*}
      \sum_{i=0}^w 
     \frac{\binom wi\binom{n-w}i}{\binom nw}  \Big( 1- \frac{ni}{w(n-w)}\Big)^r&=
           \sum_{i=0}^w 
    \frac{\binom wi\binom{n-w}{w-i}}{\binom nw}   \Big( 1- \frac{n(w-i)}{w(n-w)}\Big)^r \\
&=    \Big(\frac{n}{w(n-w)}\Big)^r  \sum_{i=0}^w 
     \frac{\binom wi\binom{n-w}{w-i}}{\binom nw}\Big(i-\frac{w^2}{n}\Big)^r.    
      \end{align*}
This completes the proof of Theorem \ref{thm:moment}.

\bibliographystyle{amsplain}
\bibliography{aryabib}

\providecommand{\bysame}{\leavevmode\hbox to3em{\hrulefill}\thinspace}
\providecommand{\MR}{\relax\ifhmode\unskip\space\fi MR }
\providecommand{\MRhref}[2]{%
  \href{http://www.ams.org/mathscinet-getitem?mr=#1}{#2}
}
\providecommand{\href}[2]{#2}
\begin{thebibliography}{10}

\bibitem{ban84}
E.~Bannai and T.~Ito, \emph{Algebraic combinatorics {I}. {A}ssociation
  schemes}, Benjamen/Cummings, London e. a., 1984.

\bibitem{bas13}
L.A. Bassalygo and V.V. Rykov, \emph{Multiple-access hyperchannel}, Problems of
  Information Transmission \textbf{49} (2013), no.~4, 399--307.

\bibitem{HCD07}
C.H. Colbourn and J.H. Dinitz (eds.), \emph{Handbook of combinatorial designs},
  2nd ed., Chapman \& Hall, 2007.

\bibitem{del73}
P.~Delsarte, \emph{An algebraic approach to the association schemes of coding
  theory}, Philips Research Repts Suppl. \textbf{10} (1973), 1--97.

\bibitem{del73a}
\bysame, \emph{Four fundamental parameters of a code and their combinatorial
  significance}, Information and Control \textbf{23} (1973), 407--438.

\bibitem{du1993combinatorial}
D.~Z. Du and F.K. Hwang, \emph{Combinatorial group testing and its
  applications}, 2nd ed., World Scientific, 2000.

\bibitem{d1982bounds}
A.~G. D'yachkov and V.~V. Rykov, \emph{Bounds on the length of disjunctive
  codes}, Problemy Peredachi Informatsii \textbf{18} (1982), no.~3, 7--13.

\bibitem{dyachkov00}
A.G. D'yachkov, A.~J. Macula, and V.V. Rykov, \emph{New applications and
  results of superimposed code theory arising from the potentialities of
  molecular biology}, Numbers, information and complexity ({B}ielefeld, 1998),
  Kluwer Acad. Publ., Boston, MA, 2000, pp.~265--282.

\bibitem{furedi96}
Z.~F{\"u}redi, \emph{On $r$-cover-free families}, Journ. Combin. Theory, Ser. A
  \textbf{73} (1996), 172--173.

\bibitem{gilbert2012recovering}
A.~C Gilbert, B.~Hemenway, A.~Rudra, M.~J. Strauss, and M.~Wootters,
  \emph{Recovering simple signals}, Information Theory and Applications
  Workshop (ITA), 2012, IEEE, 2012, pp.~382--391.

\bibitem{Hansen90}
Y.~Hansen and H.~Stichtenoth, \emph{Group codes on certain algebraic curves
  with many rational points}, Appl. Alg. Commun. Contr. Comput. \textbf{1}
  (1990), 67--77.

\bibitem{hoeffding1963probability}
W.~Hoeffding, \emph{Probability inequalities for sums of bounded random
  variables}, J. Amer. Statist. Assoc. \textbf{58} (1963), 13--30.

\bibitem{hwang1987non}
F.K. Hwang and V.~S{\'o}s, \emph{Non-adaptive hypergeometric group testing},
  Studia Sci. Math. Hungar \textbf{22} (1987), 257--263.

\bibitem{Johnson85}
W.~B. Johnson, G.~Schechtman, and J.~Zinn, \emph{Best constants in moment
  inequalities for linear combinations of independent and exchangeable random
  variables}, Ann. Probab. \textbf{13} (1985), 234--253.

\bibitem{kautz1964nonrandom}
W~Kautz and R.~Singleton, \emph{Nonrandom binary superimposed codes}, IEEE
  Trans. Inform. Theory \textbf{10} (1964), no.~4, 363--377.

\bibitem{Keevash14}
P.~Keevash, \emph{The existence of designs}, arXiv preprint arXiv:1401.3665
  (2014).

\bibitem{kuperberg2013probabilistic}
G.~Kuperberg, S.~Lovett, and R.~Peled, \emph{Probabilistic existence of regular
  combinatorial structures}, arXiv:1302.4295 (2013).

\bibitem{macula2004trivial}
A.~J. Macula, V.~V. Rykov, and S.~Yekhanin, \emph{Trivial two-stage group
  testing for complexes using almost disjunct matrices}, Discrete Applied
  Mathematics \textbf{137} (2004), no.~1, 97--107.

\bibitem{MS1977}
F.~J. MacWilliams and N.~J.~A. Sloane, \emph{The theory of error-correcting
  codes}, North-Holland, 1977.

\bibitem{malyutov1977mathematical}
M.~B. Malyutov, \emph{Mathematical models and results in the theory of
  screening experiments}, Problems of Cybernetics, no.~35, USSR Academy of
  Sciences, Moscow, 1977, in Russian, pp.~5--69.

\bibitem{malyutov1978separating}
\bysame, \emph{The separating property of random matrices}, Mathematical Notes
  \textbf{23} (1978), no.~1, 84--91.

\bibitem{mazumdar2012almost}
A.~Mazumdar, \emph{On almost disjunct matrices for group testing}, Algorithms
  and Computation, Springer, 2012, pp.~649--658.

\bibitem{mazumdar2015}
\bysame, \emph{Nonadaptive group testing with random set of defectives}, IEEE
  Trans. Inform. Theory \textbf{62} (2016), no.~12, 7522--7531.

\bibitem{nguen88}
Q.~A. Nguen and T.~Zeisei, \emph{Bounds on constant weight binary superimposed
  codes}, Probl. Contr. Inform. Theory \textbf{17} (1988), 223--230.

\bibitem{Peshkir95}
G.~Peshkir and A.~N. Shiryaev, \emph{The {K}hintchine inequalities and
  martingale expanding sphere of their action}, Uspekhi Mat. Nauk \textbf{50}
  (1995), no.~5(305), 3--62, translation in Russian Math. Surveys 50 (1995),
  no. 5, 849--904.

\bibitem{pless1963power}
V.~Pless, \emph{Power moment identities on weight distributions in error
  correcting codes}, Information and Control \textbf{6} (1963), no.~2,
  147--152.

\bibitem{porat2008explicit}
E.~Porat and A.~Rothschild, \emph{Explicit non-adaptive combinatorial group
  testing schemes}, Automata, languages and programming. {P}art {I}, Lecture
  Notes in Comput. Sci., vol. 5125, Springer, Berlin, 2008, pp.~748--759.

\bibitem{ren2001best}
Y.-F. Ren and H.-Y. Liang, \emph{On the best constant in
  {M}arcinkiewicz--{Z}ygmund inequality}, Statistics \& Probability Letters
  \textbf{53} (2001), no.~3, 227--233.

\bibitem{Rosenthal70}
H.P. Rosenthal, \emph{On the subspaces of $l^p (p>2)$ spanned by sequences of
  independent random variables}, Israel J. Math. \textbf{8} (1970), no.~3,
  273--303.

\bibitem{ruszinko94}
M.~Ruszink{\'o}, \emph{On the upper bound on the size of the $r$-cover-free
  families}, Journal of Combinatorial Theory, Series A \textbf{66} (1994),
  302--310.

\bibitem{sid75}
V.~M. Sidelnikov, \emph{Upper bounds on the cardinality of a binary code with a
  given minimum distance}, Information and Control \textbf{28} (1975), no.~4,
  292--303, Translated from the Russian by A. M. Odlyzko (Problemy Pereda\v ci
  Informacii {\bf 10} (1974), no. 2, 43--51).

\bibitem{sid09}
V.~M. Sidelnikov and O.~Yu. Prikhodov, \emph{On the construction of {$(w,r)$}
  cover-free codes}, Probl. Inform. Trans. \textbf{45} (2009), no.~1, 36--40.

\bibitem{sti93}
H.~Stichtenoth, \emph{Algebraic function fields and codes}, 2nd ed.,
  Springer-Verlag, Berlin, 2009.

\bibitem{stinson2004combinatorial}
D.~R. Stinson, \emph{Combinatorial designs: constructions and analysis},
  Springer, New York, 2004.

\bibitem{tar82}
H.~Tarnanen, \emph{An approach to constant-weight and {L}ee codes by using the
  methods of association schemes}, Annals Univ. Turkuensis, Ser. A, vol. 182,
  Turku, 1982.

\bibitem{UMB16}
S.~Ubaru, A.~Mazumdar, and A.~Barg, \emph{Group testing schemes from low-weight
  ccodeword of {BCH} codes}, Proc. 2016 IEEE Int. Sympos. Inform. Theory
  (ISIT), Barcelona (Spain), 2016, pp.~2863--2867.

\bibitem{zhigljavsky2003probabilistic}
A.~Zhigljavsky, \emph{Probabilistic existence theorems in group testing},
  Journal of Statistical Planning and Inference \textbf{115} (2003), no.~1,
  1--43.

\end{thebibliography}

\end{document}